\documentclass{aastex}          
\usepackage{spr-astr-addons}    
\usepackage{url}

\begin{document}
%
\title{Asymptotic Giant Branch Variables in the Galaxy and the Local Group}

\shorttitle{AGB Variables}
\shortauthors{Patricia Whitelock}

\author{Patricia A Whitelock} 

\altaffiltext{1}{South African Astronomical Observatory, P O Box 9,
Observatory, 7935, South Africa and}
\altaffiltext{2}{Astrophysics, Cosmology and Gravity Centre, Astronomy
Department, University of Cape Town, Rondebosch, 7700, South Africa.}

\begin{abstract} AGB variables, particularly the large amplitude Mira type,
are a vital step on the distance scale ladder.  They will prove particularly
important in the era of space telescopes and extremely large ground-based
telescopes with adaptive optics, which will be optimized for infrared
observing.  Our current understanding of the distances to these stars is
reviewed with particular emphasis on improvements that came from Hipparcos
as well as on recent work on Local Group galaxies.  In addition to providing
the essential calibration for extragalactic distances Gaia may also provide
unprecedented insight into the poorly understood mass-loss process itself. 
\end{abstract}


%
\section{Introduction}
 I will be discussing Asymptotic Giant Branch (AGB) variables, with a strong
emphasis on the pulsating stars with the largest amplitudes: the Mira
variables.  My main intention is to convince you that these stars offer a
huge potential as distance calibrators for distant resolved stellar
populations.  Under many circumstances they will be a great deal better than
Cepheids, but if they are to be used effectively then it is essential that
we use the opportunity offered by Gaia to calibrate them properly.  The very
large, but variable, size together with significant spatial asymmetries
expected for these stars will offer some special challenges in the
interpretation of the information we will get from Gaia.

Mass loss, and indeed most aspects of AGB phase evolution, remain
poorly understood. Our
interest in Gaia and AGB stars is therefore two-fold: (1) there is empirical
evidence, as discussed below, that Mira variables are good distance
indicators and we require the Gaia calibration of local Miras to put this to
use in distant galaxies; (2) the detailed observations by Gaia of
atmospheric and dust shell structures of Miras will potentially provide
invaluable insight into the mass-loss process and into AGB evolution. 

\begin{figure}
\includegraphics[width=\columnwidth]{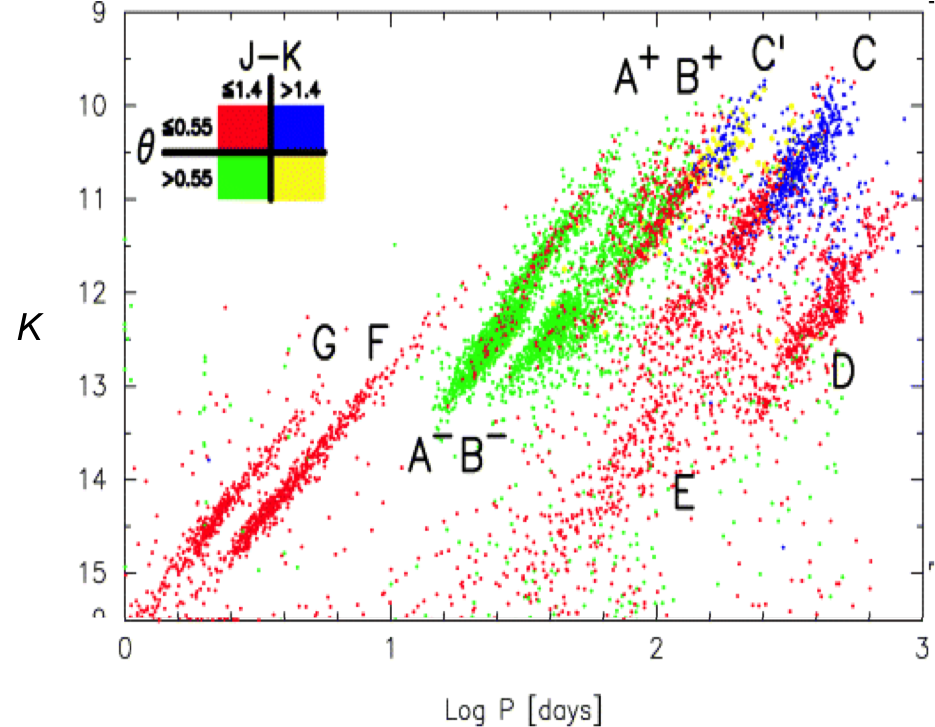}
\caption{PL($K$) relation for the LMC from \citet{Ita2004}} 
\label{fig:ita}
\end{figure}

\section{Mira Variables and their Evolutionary Status}
 Mira variables are distinguished by their large pulsation amplitudes
($\Delta I > 0.8$ or 0.9, $\Delta K > 0.4$ mag).  They are late-type stars
with spectra that are dominated by molecular absorption features in the
optical and near-infrared (e.g.  TiO, ZrO, CN, $\rm C_2$, $\rm H_2O$) and by
emission and/or absorption from dust grains (largely silicate or
carbonaceous, depending on the C/O ratio) in the mid-infrared.  Shock waves
associated with the large amplitude pulsations result in emission lines at
certain phases and, where the visible spectra can be observed, they are Me
or Ce-type, or more rarely Se or Ke.  Pulsation periods range from just
under 100 days to over 2000 days, although by far the majority are in the
200 to 450 day range \citep{samus09}.

Stars in the approximate mass range $0.8-8M_{\odot}$ evolve through the AGB
after they experience core He-burning and before planetary nebula formation. 
Somewhat more massive stars may become super-AGB stars, but their evolution
is less clear.

During the AGB most of the stellar energy originates from H-shell burning,
although the occasional He-shell flash has a profound influence on the
star, changing not only the luminosity and pulsation behaviour, but also
provoking the dredge-up of carbon from the core that is responsible for
changing the chemistry from O-rich to C-rich.

The Mira phase is relatively short lived, of the order of $10^5$ years for
low mass stars, and is the evolutionary period during which the star is at
its most luminous and experiencing its highest mass-loss rate.  In general,
as stars ascend the AGB they go through various stages of instability and
will be observed as low amplitude, overtone pulsators before they become
Miras.  Of course He-shell flashes complicate this picture and it is
possible that stars enter and leave the Mira phase more than once, although
this is not certain.  It has been demonstrated that AGB variables in the
Large Magellanic Cloud (LMC) fall on a number of period luminosity (PL)
relations (e.g.  \cite{Wood2000}; \citet{Ita2004}), as illustrated in
Fig.~\ref{fig:ita}.  



\section{Miras and Cepheids - a Comparison}
 While there are a variety of ways for estimating the distances of galaxies,
I don't really need to remind this audience that the Cepheid variables have
provided a really vital step in bridging the gap between the Local Group and
distant unresolved systems.  It is therefore interesting to examine under
what circumstances Miras might prove as useful as or more so than, Cepheids.

The next generation of telescopes, in space and on the ground, will be
operating primarily at infrared wavelengths.  So it is in this spectral
region that we should make the comparison and Table~\ref{tbl:MC} shows the
absolute $K$ and $8\mu$m magnitudes for Cepheids and Miras with various
different pulsation periods.  At $K$ the two types of variable are
comparable, but as one goes to longer wavelengths the rising relative
brightness of the Miras renders them increasingly attractive. Of course at
$ 8 \mu$m, much of the emission from Miras with P=380 days will originate
from dust \citep[e.g.][]{Riebel2010}.

As Miras belong to a somewhat older population than Cepheids they are found
in elliptical and dwarf spheroidal galaxies. Within spiral galaxies the
Miras are less concentrated in the spiral arms than are Cepheids and
therefore less crowded and more easily resolved at large distances.  

On the negative side, it could take longer to establish the period of a Mira, 
they are less regular than Cepheids, have larger amplitudes and are not as
well understood. 

 \begin{table}
\begin{center}
 \caption{A comparison of the absolute magnitudes of Cepheid and Mira
variables at 2.2 and $8\mu$m (Feast 2010 unpublished)} 
 \label{tbl:MC}
 \begin{tabular}{cccc}
 \tableline  
wavelength & variable & period & absolute \\
           &  type   & (day)  & mag \\
\tableline
$K$(2.2$\mu$m) & Cep & 50 & $-7.9$\\
               & M   & 380 & $-7.9$\\
\tableline
  8$\mu$m      & Cep & 50 & $-8.3$\\
               & M   & 230 & $-8.3$\\
               & M   & 380 & $-9.2$\\
 \tableline 
 \end{tabular}
\end{center}
 \end{table}

\section{Period-Luminosity Relations for the Distance Scale}
 We tend to work with the PL relation at $K$, PL($K$), because $K$ is
reasonably easy to measure, the amplitude of the variability is less at $K$
than at shorter wavelengths (typically $\Delta K >0.4$ mag, $\Delta I>0.9$
mag and $\Delta B>2.5$ mag), it is less affected by circumstellar
absorption than shorter wavelengths and less affected by circumstellar
emission than longer wavelengths.  However, the PL($K$) relation clearly breaks down
when there is significant circumstellar reddening.  This occurs at very long
periods for O-rich Miras, but over quite a large range of periods for C-rich
stars (see also section 6).
  
The obvious alternative to the PL($K$) relation  is the bolometric PL relation. This has the
advantage that it is independent of circumstellar reddening and that it can
be more easily compared to models, i.e. it is arguably more fundamental.
However, it is much more difficult to measure as it properly requires
observations over a large range of wavelength and over a long period of
time. There are short-cuts, for example by using colour-dependent bolometric
corrections, but problems can arise because different methods give
systematically different results \citep{Whitelock2009}.

Exactly what is done depends on what one is trying to achieve and the type
of data available, but in general $K$ remains preferable for determining
distances, provided that $A_K$ is low or measurable 
\citep[e.g.][]{Matsunaga2009}.

Figure \ref{fig:ita} shows the PL($K$) relations for giant branch and AGB
variables in the LMC.  The periods came from OGLE II and the $K$-mags from
the InfraRed Survey Facility (IRSF) at Sutherland in South Africa.  Miras,
the largest amplitude variables, are found mainly on sequence C with a very
few on the long-period part of $\rm C'$.  Because the periods come from OGLE there
is a bias towards relatively thin shelled sources that are bright in the
red.  If one looked at a group selected by IRAS or Spitzer, one would find
many more stars that are fainter at $K$.

\begin{figure}
\includegraphics[width=\columnwidth]{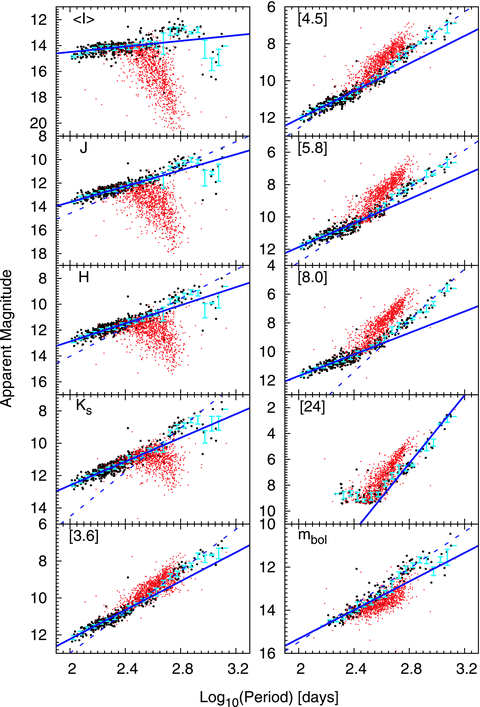}
\caption{PL relations for the LMC from \citet*{Ita2011}; 
the small red points represent presumed C-stars and the larger black
crosses presumed O-rich ones} 
\label{fig:ita2}
\end{figure}

Another example (Fig.~\ref{fig:ita2}) comes from \citet*{Ita2011}; it shows
only the large amplitude (Mira) variables with periods from OGLE~III and
uses IR data from IRSF and the Spitzer satellite.  The spectral type
distinction is made on the basis of colour, with redder stars assumed to be
C-rich.  There are two points to recognize in this figure.  First, many of
the C stars, particularly the longer period ones, fall below the PL
relation, and particularly so at shorter wavelengths.  The reason for this
is well understood: these are stars with dust shells that are sufficiently
thick that they affect the $K$-mag; the stars are red (their separation from
the PL($K$) relation will be proportional to $J-K$).  At wavelengths longer
than $3 \mu$m the thick shells emit rather than absorb, and the points fall
above the PL relations.

Secondly, at periods above about 400 days the O-rich stars show a lot of
scatter in their magnitudes, and many of them fall above the PL relation. 
This was noted earlier by e.g., \citet{Feast1989}.  The reason for this is
less clear, but it may be that the higher luminosity stars are undergoing
hot bottom burning (HBB) \citep{Whitelock2003} and therefore depart from the
PL relation, which is consequence of the core-mass luminosity relation.  There are
much fainter long-period O-rich stars, as discussed by \citet{Whitelock2003},
but these are too faint to be measured at $I$ by OGLE, they are, however,
close to the bolometric PL relation.

Note that both Figs.~\ref{fig:ita} and \ref{fig:ita2} show a good deal of
scatter because they illustrate single observations rather than mean
magnitudes; i.e., the scatter does {\it not} represent the real scatter of
the PL relation.

The behaviour of the long period stars is astrophysically very interesting,
but adds to the complexity of using them for distance determination.  It
seems sensible to suggest that we should concentrate on stars with periods
less than 400 days in distance calibration work.

\begin{figure}
\includegraphics[width=\columnwidth]{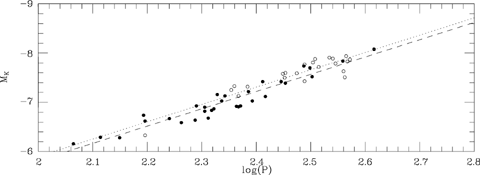}
\caption{PL relation for the LMC from \citet{Whitelock2008}; 
the solid symbols are O-rich stars, the open ones are C-rich stars} 
\label{fig:paw1}
\end{figure}

The best PL relation for distance scale work is arguably that established by
\citet*{Whitelock2008}, as illustrated in Fig.~\ref{fig:paw1}.  This is
based on multiply observed LMC sources whose status as C- or O-rich stars
was spectroscopically determined.  Very similar relations were established
for the LMC by \citet*{Ita2011} using a larger number of stars with single
$K$ observations and also for NGC\,5128 by \citet{Rejkuba2004} for
colour-selected O-rich stars with multiple observations.  This comparison
with Rejkuba also indicates that there is no significant difference in the
slope of the PL relation in the very different environments of the LMC and
NGC\,5128.
 
\citet*{Whitelock2008} also examined the PL relation zero-point calibration
for various samples of Galactic stars, assuming that the slope of the PL
relation determined for the LMC also applies to the Galaxy.  They derive the
zero-point, $\delta$, of the PL relation of the form: $$ M_K=-3.69[\log P
-2.38] + \delta $$ Their results are shown in Table~2, where the first
column gives the number of stars used to derive the zero point.

The top part of the table describes O-rich stars the lower part C-rich
stars.  The first two lines of the O-rich list, and the first of the C-rich
list, describe the results derived from the Hipparcos parallaxes.  Note that
an early paper \citep{whitelock2000} had identified short period red Miras
as a kinematically different population from their long period counterparts,
hence they were excluded from the Hipparcos results listed here.  The
selection of large amplitude stars ($\Delta H_P>1.5$ mag) from the Hipparcos
sample ensures that the stars are really Miras.  \citet*{Whitelock2008}
quoted more detailed results, but those listed here are the preferred
solutions.

The new result from \citet{Matsunaga2009} for Miras near the Galactic Centre
is also included in this table.  The Matsunaga et al.  result will give a
distance to the Galactic Centre of $(m-M)_0=14.58\pm 0.02\pm0.11 $, if we
assume, following Whitelock et al., that the distance modulus of the LMC is
18.39 (for consistency with the rest of the table).

What is seen for O-rich stars is good agreement between zero-points from
various Galactic methods including Hipparcos, globular clusters, VLBI
distances and the Galactic Centre.  While there could be metallicity
effects, it is clear that they must be small.  There is also agreement for
C-stars, but the sample is small and the uncertainties large.

%
\begin{table*}
\begin{center}
 \caption{PL($K$) relation Zero-Point ($\delta$)} 
 \begin{tabular}{cccl}
No. & $\delta$ & $\sigma_\delta$ & stars included\\
168 & --7.27 & 0.14 & Hipparcos $\pi$, $\Delta H_P> 1.5$ mag; excluding SP-red stars\\
42  & --7.32 & 0.10 & Hipparcos $\pi$, Miras + non-Miras with $\Delta H_P> 1.5$ mag;
high weight $\pi$ only\\
5   & --7.08 & 0.17 & VLBI $\pi$ for OH-Miras\\
11  & --7.34 & 0.13 & Globular cluster Miras (cluster distances from Hipparcos subdwarfs)\\
29  & --7.04 & 0.11 & Miras in Galactic Bulge $(m-M)_0=14.44$ mag\\
    & --7.18 & 0.11 & Miras near the GC \citep{Matsunaga2009}\\
31  & --7.15 & 0.06 & LMC O-rich Miras\\ 
\tableline
16  & --7.18 & 0.37 & Hipparcos $\pi$, $\Delta H_P>1.0$, hi weight C stars\\
22  & --7.24 & 0.07 & LMC C-rich Miras\\
\end{tabular}
\end{center}
\end{table*}

Figure 4 illustrates the PL($K$) relation for O-rich Galactic stars with
good individual parallax measurements.  This is taken from
\citet{Whitelock2008}, but includes the recent $\rm H_2O$ parallax
measurements made by VERA (\citet{Kurayama2005} for UX~Cyg and
\citet{Nyu2011} for SY Scl).  The high luminosity long period point, UX Cyg
(P=561), is interesting as this star may well be hot bottom burning.

\begin{figure}
\includegraphics[width=\columnwidth]{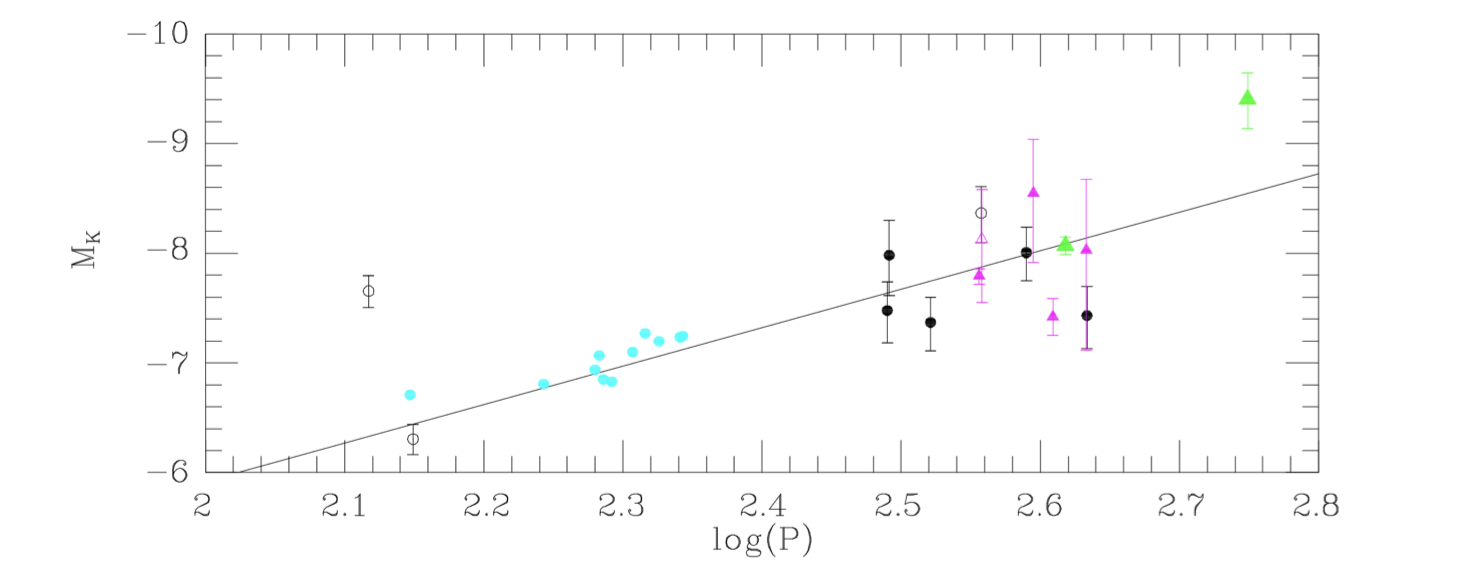}
\caption{PL relation for Galactic Miras with good distances,
from \citet{Whitelock2008}, with additional data; open symbols are SR
Variables, the others are Miras; cyan points represent globular clusters,
black circles are Hipparcos Miras with high S/N, magenta triangles are VLBI
parallaxes and green triangles are the new VLBI data from VERA} 
\label{fig:paw4}
\end{figure}

Figure 5 was prepared by Christoph Hagspihl, a student from the  South
African National Astrophysics and Space Science Programme (NASSP) as part of
his research project.  It shows the 2MASS $K$ magnitudes for Miras with
periods from OGLE\,III.  Some of the scatter will be due to the single $K$
observations, while the faint stars centred around $\log P = 2.6$ will be
the thick shelled C-stars discussed above in the context of Fig.~2. 
Superimposed on this are stars known to be hot bottom burning, from both the
LMC and SMC (assuming a distance modulus difference of 0.4 mag), including
three SR variables.  The HBB sample includes stars with strong lithium
absorption from \citet{Smith1995} and those with strong rubidium from
\citet{Garcia2009}.  The HBB stars are among the most luminous stars for
their period, but it remains possible that this is a selection effect - it
is easier to get the high signal-to-noise and good resolution required to
make the observation from luminous stars.

\begin{figure}
\includegraphics[width=\columnwidth]{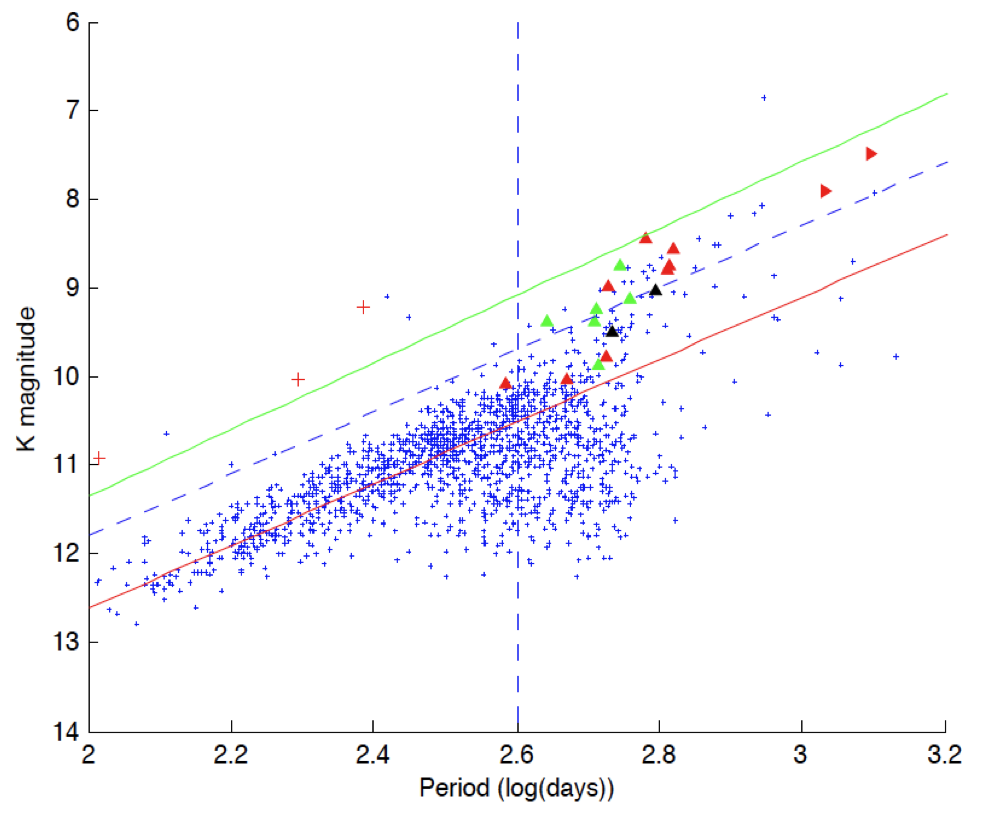}
\caption{PL relation for LMC Miras (3 crosses represent SR variables) 
showing HBB stars from LMC and SMC, compared to other Miras. LMC and SMC
Li-rich Miras from \citet{Smith1995} are plotted as upright red and green 
triangles respectively; the strong HBB candidates from \citet{Whitelock2003} 
are shown as black triangles and the Rb-rich OH/IR stars from
\citet{Garcia2009} are indicated as horizontal red triangles.} 
\label{fig:paw5}
\end{figure}

In this section I have concentrated on the PL($K$) relation, as I believe
that will be the most useful for distance calibration.  There are occasions
when we need to use the bolometric PL relation, as described below.

\section{Local Group Galaxies}  

A group of us based at the South African Astronomical Observatory and the
University of Cape Town, in collaboration with colleagues in Japan, have
been observing Local Group galaxies in the near infrared, $JHK$, using the
1.4m Infrared Survey Facility (Matsunaga this volume) at SAAO Sutherland.
 
To date we have published observations for 4 dwarf spheroidals, Fornax,
Leo~I, Sculptor and Phoenix, which have 17 Miras between them
(\citet{Menzies2008}, \citet{Whitelock2009b}, \citet{Menzies2010},
\citet{Menzies2011}).  We have started work on NGC\,6822, a barred dwarf
irregular which has over 50 Miras.  Much of the work on NGC\,6822 was done
by Francois Nsengiyumva, an MSc student.

Figure~6 shows a combined PL($K$) diagram for all of the dwarf spheroidals,
calculated assuming an LMC distance modulus of 18.39 mag.  If one
changes the assumed distance to the LMC the points all move up or down in
the diagram, but their relative positions remain unchanged.  There is a good
deal of scatter in this diagram and no clearly defined PL($K$) relationship.

\begin{figure}
\includegraphics[width=\columnwidth]{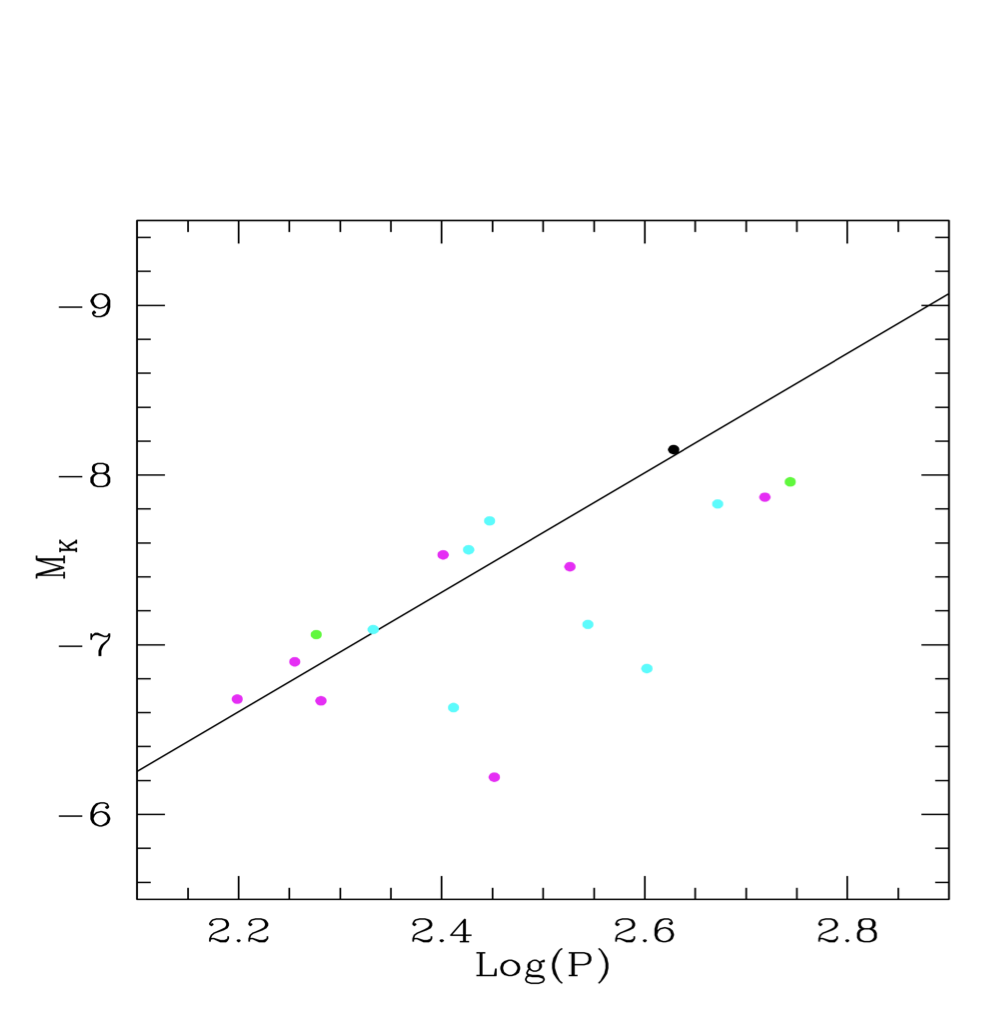}
\caption{PL($K$) relation for Local Group Dwarfs: Fornax (7), Leo\,I (7), 
Sculptor (2) and Phoenix (1)} 
\label{fig:paw6}
\end{figure}

\begin{figure}
\includegraphics[width=\columnwidth]{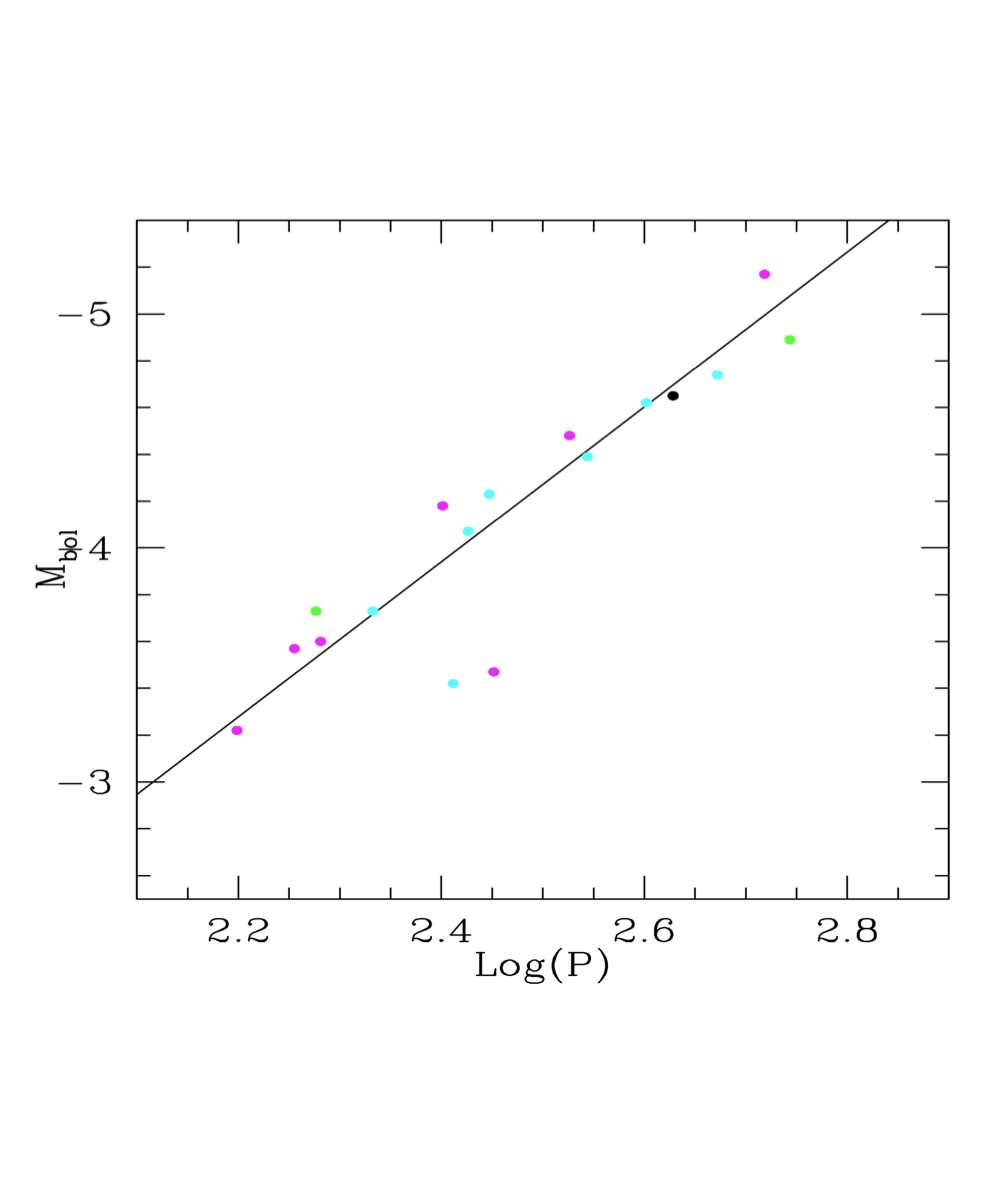}
\caption{Bolometric PL relation for the same Local Group Dwarfs as shown in
Fig.~\ref{fig:paw6}} 
\label{fig:paw7}
\end{figure}

Where we have spectra for these stars we know them to be C-rich and on the
basis of their colour we would anticipate that they are all C-rich.  On this
assumption bolometric magnitudes can be estimated using a colour-dependent
bolometric correction (Whitelock et al.  2009).  Figure~7 shows the
resulting bolometric PL relation which, with the exception of two points,
shows a tight relationship.  This demonstrates that most of the
scatter seen in Fig~6  was caused by circumstellar absorption of varying
amounts.

It seems that these Local Group C-rich Miras exhibit thick dust shells
rather more commonly than their Galactic or LMC counterparts.  There are two
possible explanations for the stars that fall below the bolometric PL
relation.  First, it may be that the bolometric correction, which was
derived for Galactic C stars \citep{Whitelock2006}, does not apply to these
short period thick shelled sources, which are not known in the Galaxy.  Note
in particular that the red side of the bolometric correction curve
(\citeauthor [fig.  15] {Whitelock2006}) is effectively a reddening locus. 
The dwarf spheroidals have some much shorter period, and presumably
intrinsically hotter, stars with thick shells.  The bolometric correction
will not adequately work for these.

Alternatively, the faint stars  are undergoing an obscuration event of the
sort described below, which are known to be quite common among
LMC \citep{Whitelock2003} and Galactic \citep{Whitelock2006} C stars.
If the obscuration is clumpy, or non-uniform over the hemisphere, then again
the bolometric correction will not give the correct bolometric magnitude. 

Figure 8 shows the bolometric PL relation for Miras and a few of the
numerous SR variables in NGC\,6822 \citep{Nsengiyumva2010}.  All of the
Miras in this diagram are thought to be C-rich, although there are also
Miras in that galaxy which are probably O-rich.  LMC Miras are shown for
comparison, again assuming a distance modulus of 18.39 mag.  This would put
NGC\,6822 at 23.43, comparable to published estimates.  The LMC and
NGC\,6822 Miras show the same slope within the uncertainties.

\begin{figure}
\includegraphics[width=\columnwidth]{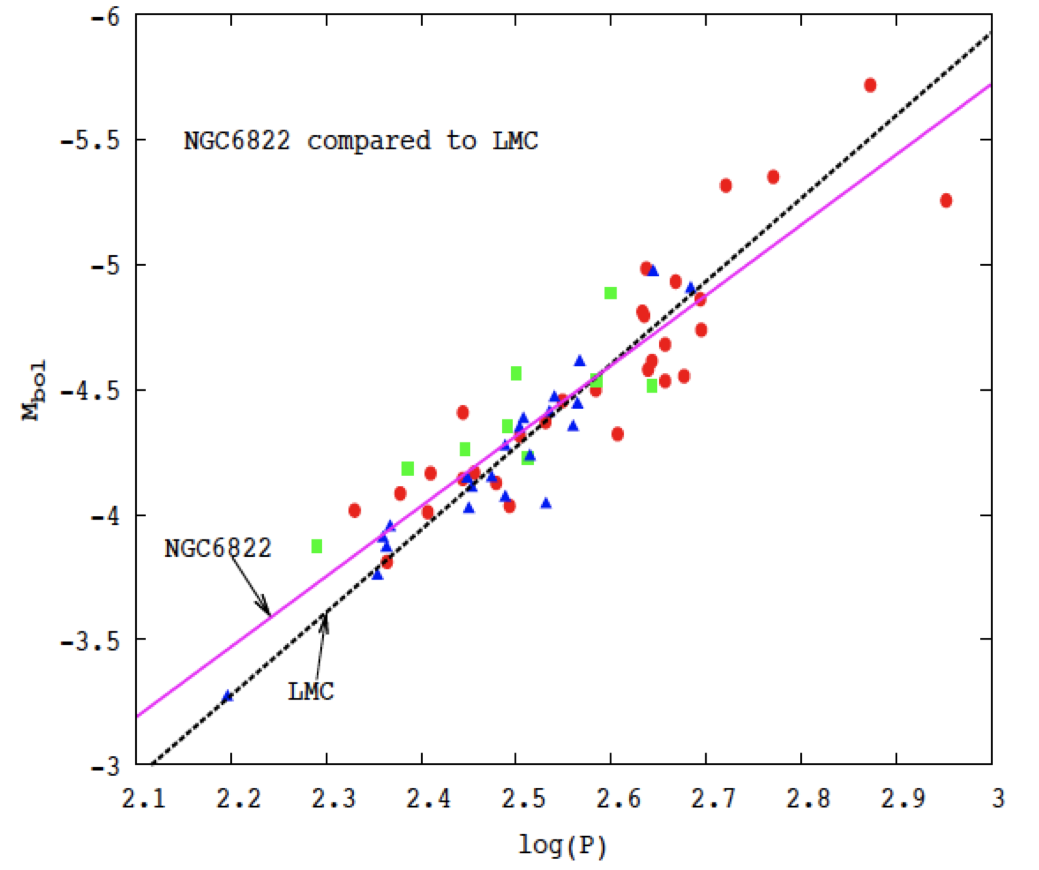}
\caption{The bolometric PL relation for NGC\,6822 from Nsengiyumva(2010); symbols 
as follows: Miras: red circles; SR variables: green squares; LMC C-rich
Miras: blue triangles \citep{Nsengiyumva2010}} 
\label{fig:paw8}
\end{figure}

\section{Obscuration Events in C-rich Stars}

\citet{Whitelock2006} noted that erratic mass-loss from C-rich Miras was
common: one third of the well studied Miras exhibited the phenomenon, as did
an unknown fraction of C-rich SR-variables. In an earlier detailed study of
the 575-day Mira, II~Lup, \citet{Feast2003} demonstrated that the mass loss
was occurring in a similar way to that from R CrB stars. This involves the
ejection of puffs of dust in random directions, rather than a uniform shell.    
It is also clear \citep{Whitelock1997} that these ejections are not
periodic.  

This phenomenon, which has only recently been recognized, will complicate
the derivation of bolometric magnitudes of these stars, as their shells
cannot be assumed to be uniform. However, the work on Local Group galaxies
discussed above suggests that a PL relation can be used to determine
distances provided that several stars can be measured for any particular
system.  

\section{Resolving circumstellar features with Gaia}
\begin{figure}
\includegraphics[width=\columnwidth]{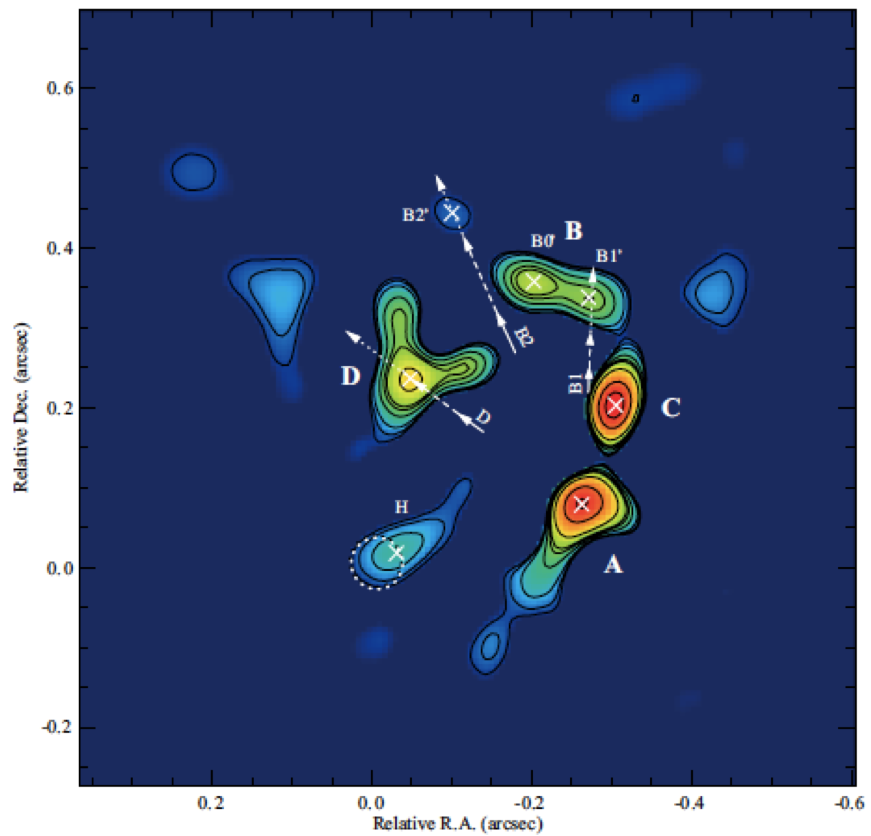}
\caption{$H$ image of the C-rich Mira CW Leo showing only the high spatial
frequencies, from \citet{Leao2006}} 
\label{fig:paw9}
\end{figure}

 I would like to finish by discussing a different perspective and by
suggesting that Gaia will potentially resolve considerable detail of the
very extended atmospheres and shells of Miras. This will be of huge interest
in its own right.  CW~Leo is one of the best-studied C-rich Miras with a
pulsation period of 650 days.  \citet{Leao2006} produced diffraction-limited
$JHKL$ images of it which showed extended structure with a central core of
clumps whose position and luminosity vary (see Fig.~\ref{fig:paw9}).  Gaia
will presumably detect each separate clump, but the interpretation as they
vary in brightness and move on the sky will be a challenge.

Mira variables are extremely large and have complex structured atmospheres. 
The problem is obvious when one recognizes that any plausible measure of the
angular diameter of a Mira will be larger than its parallax, even if they
are not all as complex as CW~Leo.  Typically the diameter of the star will
be around 1 AU and the diameter of its dust shell will be about 1000AU, so
for Gaia, Miras are not point sources.

I am not a Gaia expert, but this seems to me to pose both a challenge and an
opportunity.  The challenge in interpreting the parallax data is clear:
interpreting the Gaia data will not be as simple as it will be for smaller
stars.  However, the Gaia data will also contain a wealth of information
about the structure of the star which could potentially provide tremendous
astrophysical insight.  Interferometry from the VLT and Keck
(\citet{Mennesson2002}, \citet{Thompson2002}) has
shown that Mira disks exhibit significant departures from uniformity.  This
may be related to the way in which they lose mass; it almost certainly
varies with pulsation phase and it probably varies in other non-predictable
ways.  Can Gaia perhaps provide the critical insight into the relationship
between pulsation and mass loss?

%
\acknowledgments
I am grateful to John Menzies and Michael Feast for a critical reading of
the manuscript and for discussions of various aspects of this paper. I also
thank the National Research Foundation (NRF) for a research grant which
supported this work.



%
\bibliographystyle{spr-mp-nameyear-cnd}  
\bibliography{mybib}                

%

\end{document}